\documentclass[a4paper,UKenglish,cleveref,autoref,thm-restate,authorcolumns]{lipics-v2019}
\nolinenumbers
\usepackage[usenames,dvipsnames,svgnames,table]{xcolor} 
\usepackage{tikz}
\usetikzlibrary{calc,shadings,patterns,tikzmark}
\usepackage[normalem]{ulem}

\newcommand\HatchedCell[4][0pt]{%
  \begin{tikzpicture}[overlay,remember picture]%
    \fill[#4] ( $ (pic cs:#2) + (0,1.9ex) $ ) rectangle ( $ (pic cs:#3) + (0pt,-#1*\baselineskip-.8ex) $ );
  \end{tikzpicture}%
}%

\newcommand*{\hatched}[2]{\multicolumn{#2}{!{\hspace*{-0.4pt}\tikzmark{start#1}}c!{\tikzmark{end#1}}}{}}

\usepackage{adjustbox}
\usepackage{pgfplots}
\usetikzlibrary{spy,chains}
\pgfplotsset{compat=newest}
\usetikzlibrary{plotmarks}
\usepgfplotslibrary{groupplots}
\definecolor{lightgray}{gray}{0.9}
\usepackage{todonotes}
\usepackage{array,multirow}
\usepackage{caption}
\usepackage{mathptmx}
\usepackage{booktabs}
\usepackage{array,multirow}
\usepackage{booktabs,rotating,eqparbox,makecell,caption}
\usepackage{siunitx}
\usepackage{etoolbox}
\apptocmd{\thebibliography}{\raggedright}{}{}
\usepackage{mathtools,amssymb,mathrsfs}
\usepackage{yfonts}
\usepackage{pifont}
\usepackage{sansmath}
\newcommand{\eps}{\epsilon}
\newcommand{\astralbody}{view}

\newcommand\myenv[1]{#1}

\newcommand\pnt[1]{\mathbf{#1}}
\DeclarePairedDelimiter{\ceil}{\lceil}{\rceil}
\DeclarePairedDelimiter{\floor}{\lfloor}{\rfloor}
\renewcommand{\O}{\mathcal{O}}
\usepackage[noend]{algpseudocode}
\algnewcommand\algorithmicassert{\texttt{assert}}
\algnewcommand\Assert[1]{\State \algorithmicassert(#1)}%
\newcommand*\Let[2]{\State #1 $\gets$ #2}
\usepackage{algorithm}
\usepackage{mathptmx}
\usepackage{algorithmicx}
\usepackage{footnote}
\newcommand{\ra}[1]{\renewcommand{\arraystretch}{#1}}
\newcommand\smallO{
  \mathchoice
    {{\scriptstyle\mathcal{O}}}
    {{\scriptstyle\mathcal{O}}}
    {{\scriptscriptstyle\mathcal{O}}}
    {\scalebox{.7}{$\scriptscriptstyle\mathcal{O}$}}
}
\usepackage{tikz}
\usepackage{caption}
\usepackage{subcaption}
\usetikzlibrary{calc,arrows,decorations.pathmorphing,intersections}
\usepackage[font={footnotesize,sf},labelfont={bf},labelsep=endash]{caption}
\usepackage{sansmath}
\graphicspath{{./images/}}
\usetikzlibrary{patterns}
\usepackage{listings}
\newcommand{\classname}[1]{\begin{small}{\texttt{#1}}\end{small}}
\newcommand{\microsec}[0]{\si{\microsecond}}
\newcommand{\ext}[0]{{\texttt{ext}}}
\newcommand{\extptr}[0]{${\mathtt{ext^\dag}}$}
\newcommand{\extrrr}[0]{${\mathtt{ext^c}}$}
\newcommand{\extun}[0]{${\mathtt{ext^p}}$}
\newcommand{\nv}[0]{{\texttt{nv}}}
\newcommand{\nvlca}[0]{${\mathtt{nv^L}}$}
\newcommand{\nvsuc}[0]{${\mathtt{nv^c}}$}
\newcommand{\nvsucc}[0]{${\mathtt{nv^c}}$}
\newcommand{\wthpd}[0]{{\texttt{whp}}}
\newcommand{\sdslite}[0]{{\texttt{sdsl-lite}}}
\newcommand{\wthpdptr}[0]{${\mathtt{whp^\dag}}$}
\newcommand{\wthpdun}[0]{${\mathtt{whp^p}}$}
\newcommand{\wthpdrrr}[0]{${\mathtt{whp^{c}}}$}
\newcommand{\laski}[1]{{\texttt{#1}}}

\newcommand{\euosm}[0]   {{\texttt{eu.mst.osm}}}
\newcommand{\pathmed}[0]   {{\texttt{PM}}}
\newcommand{\pathcnt}[0]   {{\texttt{PC}}}
\newcommand{\pathrp}[0]   {{\texttt{PR}}}
\newcommand{\pathsel}[0]   {{\texttt{PS}}}
\newcommand{\pathrpt}[0]   {{\texttt{PR}}}
\newcommand{\eudimacs}[0]{{\texttt{eu.mst.dmcs}}}
\newcommand{\euemst}[0]  {{\texttt{eu.emst.dem}}}
\newcommand{\mars}[0]    {{\texttt{mrs.emst.dem}}}

\title{Path Query Data Structures in Practice}
\titlerunning{Path Query Data Structures}

\author{Meng He}{Faculty of Computer Science, Dalhousie University, Canada}{mhe@cs.dal.ca}{}{}
\author{Serikzhan Kazi}{Faculty of Computer Science, Dalhousie University, Canada}{skazi@dal.ca}{}{}

\authorrunning{M., He and S., Kazi}
\Copyright{Meng He and Serikzhan Kazi}

\funding{{This work was supported by NSERC of Canada.}}

\DeclareSIUnit{\microsecond}{\SIUnitSymbolMicro s} 

\begin{document}

\maketitle 
\begin{abstract}
    We perform experimental studies on data structures that answer
    path median, path counting, and path reporting queries
    in weighted trees.
    These query problems generalize the well-known 
    range median query problem in arrays, as well as
    the $2d$ orthogonal range counting and 
    reporting problems in planar point sets, to tree structured data.
	We propose practical realizations of the latest theoretical results
	on path queries. Our data structures, which use tree extraction,
	heavy-path decomposition and wavelet trees,
	are implemented in both succinct and pointer-based form.
   	Our succinct data structures
	are further specialized to be plain or entropy-compressed. Through experiments on large sets, 
    we show that succinct data structures for path queries
	may present a viable alternative to standard pointer-based realizations,
    in practical scenarios. Compared
    to na{\"i}ve approaches that compute the answer by explicit traversal of the query path,
    our succinct data structures are several times faster 
    in path median queries and perform comparably in path counting and path reporting queries,
    while being several times more space-efficient.
    Plain pointer-based realizations of our data structures,
    requiring a few times more space than the na{\"i}ve ones,
    yield up to $100$-times speed-up over them.
\end{abstract}


\section {Introduction}\label{section:introduction}
    Let $T$ be an ordinal tree on $n$ nodes, with each node $x$ associated with a {\textit{weight}} $\pnt{w}(x)$ over 
    an alphabet $[\sigma].$\footnote{we set $[n] \triangleq \{1,2,\ldots,n\}.$}
    A {\textit{path query}} in such a tree asks to evaluate a certain given
    function on the path $P_{x,y}$, which is the path between two given query nodes, $x$ and $y.$
    A {\textit{path median}} query asks for the median weight on $P_{x,y}.$
    A {\textit{path counting}} ({\textit{path reporting}})
    query counts (reports)
    the nodes on $P_{x,y}$ with weights falling inside the given query weight range.
    These queries generalize the range median problem on arrays,
    as well as the $2d$ orthogonal counting and reporting
    queries in point sets, by replacing one of the dimensions with tree topology.
    Formally, query arguments consist of a pair of vertices $x,y \in T$ along with an interval $Q.$
	The goal is to preprocess the tree $T$ for the following types of queries:
	\begin{itemize}
        \item {\emph{Path Counting}}: return $|\{z \in P_{x,y}\,|\,\pnt{w}(z) \in {Q}\}|$. 
        \item {\emph{Path Reporting}}: enumerate $\{z \in P_{x,y}\,|\,\pnt{w}(z) \in {Q}\}$.
        \item {\emph{Path Selection}}: return the $k^{th}$ 
        ($0 \leq k < |P_{x,y}|$) weight in the sorted list of weights on $P_{x,y};$
		$k$ is given at query time.
        In the special case of $k = \floor{|P_{x,y}|/2},$ 
        a path selection is a {\textit{path median query}}.
	\end{itemize}
    Path queries is a widely-researched topic in computer science community
	~\cite{Alon87optimalpreprocessing,
	Chazelle1987,
    DBLP:journals/iandc/Hagerup00,
	DBLP:journals/njc/KrizancMS05,
	DBLP:journals/talg/0001MZ16,
	DBLP:journals/algorithmica/ChanHMZ17,
	DBLP:conf/isaac/0001K19}.
    Apart from theoretical appeal, queries on tree topologies reflect
    the needs of efficient information retrieval from hierarchical data,
    and are gaining ground in established domains such as RDBMS~\cite{ltree}.
    The expected height of $T$ being $\sqrt{2\pi{}n}$~\cite{RenyiSzekeres1967}, this
    calls for the development of methods beyond na{\"i}ve.

    Previous work includes that of Krizanc et al.~\cite{DBLP:journals/njc/KrizancMS05}, who were the first to introduce path median query problem (henceforth \pathmed) in trees,
    and gave an $\O(\lg{n})$ query-time
    data structure with the space cost of $\O(n\lg^{2}n)$ words. They also gave an $\O(n\log_b{n})$ words
    data structure to answer \pathmed{} queries in time $\O(b\lg^{3}n/\lg{b}),$ for any fixed $1 \leq b \leq n.$
    Chazelle~\cite{Chazelle1987} gave an emulation dag-based linear-space data structure
    for solving path counting (henceforth \pathcnt) queries in trees in time $\O(\lg{n}).$

    While~\cite{DBLP:journals/njc/KrizancMS05,Chazelle1987} design different data structures for {\pathmed} and {\pathcnt},
    He et al.~\cite{DBLP:conf/isaac/HeMZ11,DBLP:journals/talg/0001MZ16} use {\textit{tree extraction}} to solve both
    {\pathcnt} and the path selection problem (henceforth \pathsel), 
    as well as the path reporting problem (henceforth \pathrp), which they were the first to introduce.
    The running times for \pathsel/\pathcnt{} were $\O(\lg\sigma),$ 
    while a \pathrp{} query is answered in $\O((1+\kappa)\lg\sigma)$ time,
    with $\kappa$ henceforth denoting output size.
    Also given is an $\O(n\lg\lg\sigma)$-words
    and $\O(\lg\sigma+\kappa\lg\lg\sigma)$ query time solution, for \pathrp{}, in the RAM model.

    Further, solutions based on {\textit{succinct}} data structures started to appear.
    (In the interests of brevity, the convention throughout this paper is that a data structure is {\textit{succinct}} if its size in bits is 
    close to the information-theoretic lower bound.)
    Patil et al.~\cite{DBLP:journals/jda/PatilST12}
    presented an $\O(\lg{n}\cdot{}\lg{\sigma})$ query time data structure for \pathsel/\pathcnt{}, occupying $6n+n\lg\sigma+\smallO(n\lg\sigma)$ bits of space.
    Therein, the tree structure and the weights distribution are decoupled and
    delegated to respectively heavy-path decomposition~\cite{Sleator:1983:DSD:61337.61338}
    and wavelet trees~\cite{DBLP:books/daglib/0038982}. Their data structure also solves \pathrp{} in $\O(\lg{n}\lg\sigma+(1+\kappa)\lg\sigma)$ query time.

    Parallel to~\cite{DBLP:journals/jda/PatilST12},
    He et al.~\cite{DBLP:conf/isaac/HeMZ11,DBLP:journals/talg/0001MZ16} devised
    a succinct data structure occupying $nH(W_T)+\smallO(n\lg\sigma)$ bits of space
    to answer \pathsel/\pathcnt{} in $\O(\frac{\lg{\sigma}}{\lg\lg{n}}+1),$
    and \pathrp{} in $\O((1+\kappa)(\frac{\lg\sigma}{\lg\log{n}}+1))$ time.
    Here, $W_T$ is the multiset of weights of the tree $T,$ and $H(W_T)$ is there entropy thereof.
    Combining tree extraction and the ball-inheritance problem~\cite{DBLP:conf/compgeom/ChanLP11},
    Chan et al.~\cite{DBLP:journals/algorithmica/ChanHMZ17}
    proposed further trade-offs,
    one of them being an $\O(n\lg^{\eps}{n})$-word structure with $\O(\lg\lg{n}+\kappa)$ query time, for \pathrp{}.

    Despite the vast body of work, little is known on the practical performance
    of the data structures for path queries,
	with empirical studies on weighted trees definitely lacking, and existing related experiments being limited to
	navigation in unlabeled trees only~\cite{DBLP:conf/alenex/ArroyueloCNS10},
	or to very specific domains~\cite{DBLP:journals/algorithms/AbeliukCN13,DBLP:journals/jea/NavarroP16}.
    By contrast, the empirical study of traditional orthogonal range queries
    have attracted much attention~\cite{DBLP:journals/tcs/ArroyueloCDDHLMNSS11,DBLP:journals/is/BrisaboaBKNS16,DBLP:conf/dcc/IshiyamaS17}.
    We therefore contribute to remedying this imbalance.

    \subsection{Our work} 
    In this article, we provide an experimental study of data structures
    for path queries. The types of queries we consider
    are \pathmed, \pathcnt, and \pathrp.
    The theoretical foundation of our work are the data
    structures and algorithms developed in~\cite{
        DBLP:conf/isaac/HeMZ11,
        DBLP:journals/jda/PatilST12,
        DBLP:journals/algorithmica/HeMZ14,
        DBLP:journals/talg/0001MZ16}.
    The succinct data structure by He et al.~\cite{DBLP:journals/talg/0001MZ16}
    is optimal both in space and time in the RAM model.
    However, it builds on components that are likely to be cumbersome in practice.
    We therefore present a practical compact implementation
    of this data structure that uses $3n\lg\sigma+\smallO(n\lg\sigma)$ bits of space
    as opposed to the original $nH(W_T)+\smallO(n\lg\sigma)$ bits of space in~\cite{DBLP:journals/talg/0001MZ16}.
    For brevity, we henceforth refer to the data structures based
    on tree extraction as \ext.
    Our implementation of \ext{} achieves the query time of $\O(\lg{\sigma})$
    for \pathmed{} and \pathcnt{} queries,
    and $\O((1+\kappa)\lg\sigma)$ time for \pathrp.
    Further, we present an exact implementation
    of the data structure (henceforth \wthpd) by Patil et al.~\cite{DBLP:journals/jda/PatilST12}.
    The theoretical guarantees of {\wthpd} are $6n+n\lg\sigma+\smallO(n\lg\sigma)$
    bits of space, with $\O(\lg{n}\lg\sigma)$ and $\O(\lg{n}\lg\sigma+(1+\kappa)\lg\sigma)$
    query times for respectively \pathmed{}/\pathcnt{} and \pathrp{}.
    Although \wthpd{} is optimal neither in space nor in time,
    it proves competitive with \ext{} on the practical datasets we use.
    Further, we evaluate time- and space-impact
    of succinctness by realizing plain pointer-based versions of both \ext{} and \wthpd{}.
    We show that succinct data structures based on {\ext} and {\wthpd} offer
    an attractive alternative for their fast but space-consuming counterparts, 
    with query-time slow-down of $30$-$40$ times
    yet commensurate savings in space.
    We also implement, in pointer-based and succinct variations,
    a na{\"i}ve approach of not preprocessing the tree at all but rather
    answering the query by explicit scanning. The succinct solutions
    compare favourably to the na{\"i}ve ones, the slowest former 
    being $7$-$8$ times faster than na{\"i}ve \pathmed, while occupying
    up to $20$ times less space. \myenv{We also compare the performance of different
    succinct solutions relative to each other.}


\section{Preliminaries}\label{section:preliminaries} \label{section:dcc:ptr}\label{section:dcc:extptr}
This section introduces notation and main algorithmic techniques at the core of our data structures.

\subparagraph*{Notation.} The $i^{th}$ node visited during a preorder traversal of the given tree $T$ is said to have {\textit{preorder rank}} $i$. 
We identify a node by its preorder rank.
For a node $x \in T,$ its set of ancestors $\mathcal{A}(x)$ includes $x$ itself.
Given nodes $x,y \in T,$ where $y \in \mathcal{A}(x)$, we set $A_{x,y} \triangleq P_{x,y}\setminus{}\{y\};$
one then has $P_{x,y} = A_{x,z}\sqcup{}\{z\}\sqcup{}A_{y,z},$ where $z = LCA(x,y).$ 
The primitives {\texttt{rank/select/access}} are defined in a standard way, i.e. ${\mathtt{rank_1}(B,i)}$
is the number of $1$-bits in positions less than $i,$ ${\mathtt{select_{1}}(B,j)}$
returns the position of the $j^{th}$ $1$-bit, and $\mathtt{access(B,i)}$
returns the bit at the $i^{th}$ position, all with respect to a given bitmap $B$,
which is omitted when the context is clear.
\begin{figure}[b]
    \includegraphics[width=\textwidth]{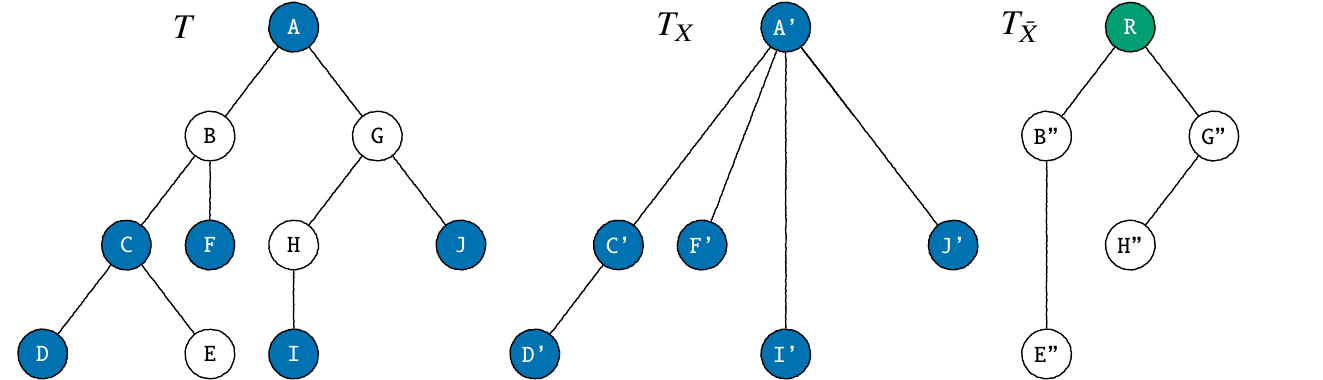}
\caption{Tree extraction. Original tree (left), extracted tree $T_X$ (middle), and extraction of the complement of $X,$ tree $T_{\bar{X}}$ (right).
     The blue shaded nodes in $T$ form the set $X.$ In the tree $T_{X},$ node $\mathtt{C'}$ 
     corresponds to node $\mathtt{C}$ in the original tree $T$, and node $\mathtt{C'}$ in the extracted tree $T_{X}$ is the $T_X$-{\astralbody} of nodes $\mathtt{C}$ and $\mathtt{E}$ in the original tree $T.$
     Finally, node $\mathtt{C}$ in $T$ is the $T$-source of the node $\mathtt{C'}$ in $T_X.$
     Extraction of the complement, $T_{\bar{X}},$ demonstrates the case of adding a dummy root $\mathtt{R}.$}
\label{figure:treeExtractionExample}
\end{figure}

\subparagraph*{Compact representations of ordinal trees.} Compact representations of ordinal trees is a well-researched area, mainstream methodologies including 
{\textit{balanced parentheses}} (BP)~\cite{DBLP:conf/focs/Jacobson89,DBLP:journals/siamcomp/MunroR01,DBLP:journals/tcs/GearyRRR06,DBLP:journals/talg/LuY08,DBLP:journals/tcs/MunroRRR12},
{\textit{depth-first unary degree sequence}} (DFUDS)~\cite{
    DBLP:journals/algorithmica/BenoitDMRRR05,
    DBLP:journals/talg/GearyRR06,
    DBLP:journals/jcss/JanssonSS12},
{\textit{level-order unary degree sequence}} (LOUDS)~\cite{
    DBLP:conf/focs/Jacobson89,
    DBLP:conf/wea/DelprattRR06}, and
{\textit{tree covering}} (TC)~\cite{
    DBLP:journals/talg/GearyRR06,
    DBLP:journals/talg/HeMS12,
    DBLP:journals/algorithmica/FarzanM14}.
Of these, BP-based representations ``combine
good time- and space-performance with rich functionality'' in practice~\cite{DBLP:conf/alenex/ArroyueloCNS10},
and we use BP in our solutions.
BP is a way of linearising the tree by emitting 
{``\texttt{(}''} upon first entering a node and {``\texttt{)}''} upon exiting, 
having explored all its descendants
during the preorder traversal of the tree.
For example, {\texttt{(((()())())((())()))}} would be a BP-sequence for the tree $T$ in \Cref{figure:treeExtractionExample}.

As shown in~\cite{DBLP:journals/siamcomp/MunroR01,DBLP:journals/talg/LuY08,DBLP:journals/tcs/MunroRRR12},
an ordinal tree $T$ on $n$ nodes can be represented in $2n+\smallO(n)$ bits of space to support the
following operations in $\O(1)$ time, for any node $x \in T:$ 
{$\mathtt{child}(T,x,i)$}, the $i$-th child of $x$;
{$\mathtt{depth}(T,x)$}, the number of ancestors of $x$; 
{$\mathtt{LCA}(T,x,y)$}, the lowest common ancestor of nodes $x,y \in T;$ and
{$\mathtt{level\_anc}(T,x,i)$}, the $i^{th}$ lowest ancestor of $x.$
\subparagraph*{Tree extraction.} Tree extraction~\cite{DBLP:journals/talg/0001MZ16} selects a subset $X$ of nodes
 while maintaining the underlying hierarchical relationship among the nodes in $X$.
 Given a subset $X$ of tree nodes called {\textit{extracted nodes}}, 
 an {\textit{extracted tree}} $T_X$ can be obtained from the original tree $T$ through the following procedure.
 Let $v \notin X$ be an arbitrary node. The node $v$ and all its incident edges in $T$ are removed from $T,$ thereby
 exposing the parent $p$ of $v$ and $v'$s children, $v_1,v_2,\ldots,v_k.$ 
 Then the nodes $v_1,v_2,\ldots,v_k$ (in this order) become new children of $p,$ 
 occupying the contiguous segment of positions starting from the (old) position of $v.$
 After thus removing all the nodes $v \notin X,$ we have $T_X \equiv F_X,$ 
 if the forest $F_X$ obtained is a tree; otherwise, a dummy root $r$ holds the roots of the trees in $F_X$ (in the original left-to-right order)
 as its children. (The symmetry between $X$ and $\bar{X} = V\setminus{}X$ 
 brings about the {\textit{complement}} $T_{\bar{X}}$ of the extracted tree $T_{X}.$)
 An original node $x \in X$ of $T$ and its copy, $x'$, in $T_X$ are said to {\textit{correspond}} to each other;
 also, $x'$ is the {\textit{$T_X$-{\astralbody}}} of $x,$ and $x$ is the {\textit{T-source}} of $x'.$
 The $T_X$-{\astralbody} of a node $y \in T$ ($y$ is not necessarily in $X$) is generally defined to be the node 
 $y' \in T_X$ corresponding to the lowest node in $\mathcal{A}(y) \cap{} X.$ 
    In this paper, tree extraction is predominantly used to classify nodes into categories,
    and the labels assigned indicate the weight ranges the original weights belong to.

 \Cref{figure:treeExtractionExample} gives an example of an extracted tree, {\astralbody}s and sources.

\section{Data Structures for Path Queries}\label{sec:compactRepr}
This section gives the design details of the {\wthpd} and {\ext} data structures.

\subsection{Data structures based on heavy-path decomposition}\label{sec:HPD}
We now describe the approach of~\cite{DBLP:journals/jda/PatilST12}, which is based on heavy-path decomposition~\cite{Sleator:1983:DSD:61337.61338}.

Heavy-path decomposition (HPD) imposes a structure on a tree.
In HPD, for each non-leaf node, a {\textit{heavy child}} is defined as the child whose subtree has the maximum cardinality.
    HPD of a tree $T$ with root $r$ is a collection of disjoint chains, first of which is obtained by always following
    the heavy child, starting from $r,$ until reaching a leaf. 
    The subsequent chains are obtained by the same procedure, starting from the non-visited nodes closest to the root (ties broken
    arbitrarily).
The crucial property is that any root-to-leaf path in the tree encounters $\O(\lg{n})$ distinct chains.
A chain's {\textit{head}} is the node of the chain that is closest to the root;
a chain's tail is therefore a leaf.

Patil et al.~\cite{DBLP:journals/jda/PatilST12} used HPD to decompose a path query
into $\O(\lg{n})$ queries in sequences. To save space, they designed the following
data structure to represent the tree and its HPD.
If $x$ is the head of a chain $\phi$, all the nodes in $\phi$ 
have a (conceptual) {\textit{reference}} pointing to $x,$ while $x$ points to itself. 
A {\textit{reference count}}
of a node $x$ (denoted as ${rc_{x}}$) stands for the number of times a node serves as a reference. 
Obviously, only heads feature non-zero reference counts -- precisely the lengths of their respective chain.
The reference counts of all the nodes are stored in unary in preorder in a bitmap $B = 10^{rc_1}10^{rc_2}\ldots{}10^{rc_{n}}$ using $2n+\smallO(n)$ bits.
Then, one has that $rc_{x} = \mathtt{rank_{0}(B,select_1(B,x+1))-rank_{0}(B,select_1(B,x))}.$ 
The topology of the original tree $T$ is represented succinctly in another $2n+\smallO(n)$ bits. In addition, they encode
the HPD structure of $T$ using a new tree $T'$ that is obtained from $T$ via the following transformation. 
All the non-head nodes become leaves and are directly connected
to their respective heads; the heads themselves (except the root) become children of the references of their original parents.
All these connections are established respecting the preorder ranks of the nodes in the original tree $T.$ Namely,
a node farther from the head attaches to it only after the higher-residing nodes of the chain have done so.
This transformation preserves the original preorder ranks. 
On $T',$ operation {\texttt{ref(x)}} is supported, which returns the head of chain to which the node $x$ in the original tree belongs.

To encode weights they call $C_x$ the weight-list of $x$ if it collects,
in preorder, all the nodes for which $x$ is a reference. Thus, a non-head node's list is empty; a head's list
spells the weights in the relevant chain. Define $C = C_1C_2\ldots{}C_{n}.$ 
Then, in $C,$ the weight of $x \in T$ resides at position
\begin{equation}\label{eq:eqnHpd}
    \mathtt{1+select_1(B,ref(x))-ref(x) + depth(x) - depth(ref(x))}
\end{equation}
(where $\mathtt{depth(x)}$ and $\mathtt{ref(x)}$ are provided by $T$ and $T',$ respectively).
\myenv{$C$ is then encoded in a wavelet tree (WT).}
To answer a query, $T,$ $T',$ $B,$ and \Cref{eq:eqnHpd} are used to partition the query
path into $\O(\lg{n})$ sub-chains that it overlaps in HPD; and for each sub-chain,
one computes the interval in $C$ storing the weights of the nodes in the chain. 
$I_m$ denotes the set of intervals computed.
    Precisely, for a node $x$, one uses $B$ to find out whether $x$ is the head of its chain; 
    if not, the parent of $x$ in $T'$ returns one (say, $y$).
    Then {\Cref{eq:eqnHpd}} maps the path $A_{x,y}$ to its corresponding interval in $C.$
    One proceeds to the next chain by fetching the (original) parent of $y$, using $T.$
Then, the WT is queried with $\O(\lg{}n)$ simultaneous (i) range quantile (for \pathmed);
or (ii) orthogonal range $2d$ queries (for \pathcnt{} and \pathrp). 

Range quantile query over a collection of ranges 
is accomplished via a straightforward extension of the algorithm of Gagie et al.~\cite{DBLP:conf/spire/GagiePT09}.
One descends the wavelet tree $W_C$ maintaining a set of current weights $[a,b]$ (initially $[\sigma]$),
the current node $v$ (initially the root of $W_C$), and $I_m.$
When querying the current node $v$ of $W_C$ with an interval $[l_j,r_j] \in I_m,$ one finds out, in $\O(1)$ time, how many weights
in the interval are lighter than the mid-point $c$ of $[a,b],$ 
and how many of them are heavier.
The sum of these values then
determines which subtree of $W_C$ to descend to.
There being $\O(\lg\sigma)$ levels in $W_C$,
and spending $\O(1)$ time for each segment in $I_m,$ the overall running time is $\O(\lg{n}\lg\sigma).$
\pathcnt/\pathrpt{} proceed by querying each interval, independently of the others, with the standard $2d$ search
over $W_C$. 

\subsection{Data structures based on tree extraction}\label{par:threelog}

The solution by He et al.~\cite{DBLP:journals/talg/0001MZ16} is based on performing
a hierarchy of tree extractions, as follows. One starts with the original tree $T$ weighted over $[\sigma]$, 
and extracts two trees $T_0 = T_{1,m}$ and $T_1 = T_{m+1,\sigma},$
respectively associated with the intervals $I_0 = [1,m]$ and $I_1 = [m+1,\sigma],$ where $m = \floor{\frac{1+\sigma}{2}}.$
Then both $T_0$ and $T_1$ are subject to the same procedure, stopping only when the current tree
is weight-homogeneous. 
We refer to the tree we have started with as the {\textit{outermost}} tree.

The key insight of tree extraction is that the number of nodes $n'$ with weights from $I_0$ on the path from $u$
to $v$ equals
$
    \mathtt{n' = depth_0(u_0) + depth_0(v_0) - 2\cdot{}depth_0(z_0) + \mathbf{1}_{w(z) \in I_0}},
$
where $\mathtt{depth_0(\cdot{})}$ is the depth function in $T_0,$ $z= LCA(u,v),$ 
$u_0,v_0,z_0$ are the $T_0$-{\astralbody}s of $u,v,$ and $z,$ 
and $\mathbf{1}_{pred}$ is $1$ if predicate $pred$ is {\sc{true}},
and $0$ otherwise. 
The key step is then, for a given node $x$, how to efficiently find its $0/1$-{\textit{parent}},
whose purpose is analogous to a {\texttt{rank}}-query when descending down the WT. 
    Consider a node $x \in T$ and its ${T_0}$-{\astralbody} $x_0.$
The corresponding node $x' \in T$ of $x_0 \in T_0$ is then called
{$0$}-{\textit{parent}} of $x.$ The ${1}$-parent is defined analogously.
Supporting $0/1$-parents in compact space is one of the main implementation challenges
of the technique, as storing the \astralbody{}s explicitly is space-expensive.
In~\cite{DBLP:journals/talg/0001MZ16}, the hierarchy of extractions is done
by dividing the range not to $2$ but $f = \O(\lg^{\eps}{n})$ parts,
with $0 < \eps < 1$ being a constant. They classify the nodes according to weights
using these $f= \ceil{\lg^{\eps}{n}}$ labels and use tree covering to represent
the tree with small labels
in order to find $T_{\alpha}$-{\astralbody}s for arbitrary $\alpha \in [\sigma],$
in constant time. They also use this representation to identify, in constant time,
which extractions to explore. Therefore, at each of the $\O(\lg{\sigma}/\lg\lg{n})$ levels
of the hierarchy of extractions, constant time work is done, yielding 
an $\O(\lg{\sigma}/\lg\lg{n})$-time algorithm for {\pathcnt}. Space-wise,
it is shown that each of the $\O(\lg\sigma/\lg\lg{n})$ levels
can be stored in $2n+nH_0(W)+\smallO(n\lg\sigma)$ bits of space in total (where $W$ is the multiset of weights on the level)
which, summed over all the levels, yields $nH_0(W_T)+\O(n\lg\sigma/\lg\lg{n})$ bits of space.
The components of this optimal result, however, use word-parallel techniques that are unlikely
to be practical. In addition, one of the components, tree covering (TC) for trees labeled over 
$[\sigma],\,\sigma= \O(\lg^{\eps}{n})$ 
has not been implemented and experimentally evaluated even for unlabeled versions thereof.
Finally, lookup tables for the word-RAM data structures may either be rendered 
too heavy by word alignment, or too slow by the concomitant arithmetic for accessing its entries.
In practice, small blocks of data are usually explicitly scanned~\cite{DBLP:conf/alenex/ArroyueloCNS10}.
However, we can see no fast way to scan small labeled trees.
At the same time, a generic multi-parentheses approach~\cite{DBLP:books/daglib/0038982}
would spare the effort altogether, immediately yielding a $4n\lg\sigma+2n+\smallO(n\lg\sigma)$-bit encoding of the tree, with $\O(1)$-time support for $0/1$-parents.
We achieve instead $3n\lg\sigma+\smallO(n\lg\sigma)$ bits of space, as we proceed to describe next. 

We store $2n+\smallO(n)$ bits as a regular BP-structure $S$
of the original tree, in which a 1-bit represents an opening
parenthesis, and a 0-bit represents a closing one, and mark in a separate length-$n$ bitmap $B$ the {\textit{types}} (i.e.~whether it is
a $0$- or $1$-node) of the $n$ opening parentheses in $S.$
The type of an opening parenthesis at position $i$ in $S$ is thus given by $\mathtt{access(B,rank_1(S,i))}.$
Given $S$ and $B$, we find the $t \in \{0,1\}$-parent of $v$ with an approach described in~\cite{DBLP:journals/algorithmica/HeMZ14}.
For completeness, we outline in \Cref{algo:algo02} how to locate the $T_t$-\astralbody{} of a node $v.$

\begin{algorithm}
    \caption{Locate the \astralbody{} of $v\in T$ in $T_t$, where $T_t$ is the extraction from $T$ of the $t$-nodes}
	\label{algo:algo02}
        \begin{algorithmic}[1]
        \Require{$t \in \{0,1\}$}
		\Function{\astralbody\_of}{$v,t$}
		\If{$\mathtt{B[v] == t}$} \Comment{$v$ is a $t$-node itself}
			\State \Return {$\mathtt{B.rank_t(v)}$} 
		\EndIf
		\Let{$\mathtt{\lambda}$}{$\mathtt{rank_t(B,v)}$} \label{alg:algo02:line:ref01} \Comment{how many $t$-nodes precede $v$?} 
		\If{$\mathtt{\lambda == 0}$} \label{alg:algo02:line:ref02} 
			\State \Return{\texttt{null}}
		\EndIf
		\Let{$\mathtt{u}$}{$\mathtt{select_t(B,\lambda)}$} \label{alg:algo02:line:ref03} \Comment{find the $\lambda^{th}$ $t$-node}
		\If{$\mathtt{LCA(u,v) == u}$}
			\State\Return{$\mathtt{B.rank_t(u)}$} \label{alg:algo02:line:ref04} 
		\EndIf
		\Let{$\mathtt{z}$}{$\mathtt{LCA(u,v)}$} \Comment{$z$ is $LCA$ of a $t$-node $u$ and a non-$t$-node $v$}
		\If{$\mathtt{z == null}$ {\bf{or}} $\mathtt{B[z] == t}$} \Comment{$z$ is a $t$-node $\Rightarrow \nexists$ $t$-parent closer to $v$}
			\State \Return $\mathtt{B.rank_t(z)}$ \Comment{or \texttt{null}}
		\EndIf
		\Let{$\mathtt{\lambda}$}{$\mathtt{rank_t(B,z)}$} \Comment{how many $t$-nodes precede $z$?}
		\Let{$\mathtt{r}$}{$\mathtt{select_t(B,\lambda+1)}$} \label{alg:algo02:line:ref05} \Comment{the first $t$-descendant of $z$}
            \Let{$\mathtt{z_t}$}{$\mathtt{rank_t(B,r)}$} \label{alg:algo02:line:ref06} \Comment{$z_t$ is the $T_t$-\astralbody{} of $r$}
			\Let{$\mathtt{p}$}{$\mathtt{T_t.parent(z_t)}$} \label{alg:algo02:line:ref07} \Comment{$p$ can be {\texttt{null}} if $z_t$ is $0$}
		\State\Return{$\mathtt{p}$}
		\EndFunction
	\end{algorithmic}
\end{algorithm}

First, find the number of $t$-nodes preceding $v$ (line~\ref{alg:algo02:line:ref01}).
If none exists (line~\ref{alg:algo02:line:ref02}), we are done; otherwise, let $u$ be the $t$-node immediately preceding $v$ (line~\ref{alg:algo02:line:ref03}).
If $u$ is an ancestor of $v,$ it is the answer (line~\ref{alg:algo02:line:ref04});
else, set $z= LCA(u,v).$
If $z$ is a $t$-node, or non-existent (because the tree is actually a forest), then return $z$ or {\texttt{null}}, respectively.
Otherwise ($z$ exists and not a $t$-node), in line~\ref{alg:algo02:line:ref05} 
we find the first $t$-descendant $r$ of $z$ (it exists because of $u$).
This descendant cannot be a parent of $v,$ since otherwise we would have found it before. 
It must share though the same $t$-parent with $v.$
We map this descendant to a node $z_t$ in $T_t$ (line~\ref{alg:algo02:line:ref06}). 
Finally, we find the parent of $z_t$ in $T_t$ (line~\ref{alg:algo02:line:ref07}).

The combined cost of $S$ and $B$ is $2n+n+\smallO(n) = 3n+\smallO(n)$ bits. At each of the $\lg\sigma$ levels
of extraction, we encode $0/1$-labeled trees in the same way, so the total space
is $3n\lg\sigma+\smallO(n\lg\sigma)$ bits.

Query algorithms in the {\ext} data structure proceed within the generic framework of extracting $T_0$ and $T_1.$
Let $n' = |P_{u_0,v_0}|.$
In \pathmed, we recurse on $T_0$ if $k < n',$ for a query that asks for a node with the $k^{th}$ smallest weight on the path $P_{u_0,v_0};$
otherwise, we recurse on $T_1$ with $k \leftarrow k-n'$ and $u_1,\,v_1.$ 
We stop upon encountering a tree with homogeneous weights. This logic is embodied in \Cref{algo:algo00}
in \Cref{appendix:queryAlgos}. Theoretical running time is $\O(\lg\sigma)$, as all the primitives used are $\O(1)$-time.

A procedure for the \pathcnt{} and \pathrpt{} is essentially similar to that for the \pathmed{} problem.
We maintain two nodes, $u$ and $v,$ as the query nodes with respect to the current extraction $T$,
and a node $z$ as the lowest common ancestor of $u$ and $v$ in the current tree $T.$
Initially, $u,v \in T$ are the original query nodes, and $T$ is the outermost tree.
Correspondingly, $z$ is the LCA of the nodes $u$ and $v$ in the original tree;
we determine the weight of $z$ and store it in $w,$ which is passed down
the recursion. Let $[a,b]$ be the query interval, and $[p,q]$ be the current
range of weights of the tree. Initially, $[p,q] = [\sigma].$
First, we check whether the current interval $[p,q]$ is contained within $[a,b].$
If so, the entire path $A_{u,z} \cup{} A_{v,z}$ belongs to the answer.
Here, we also check whether $w \in [a,b].$
Then we recurse on $T_t$ ($t \in \{0,1\}$) having computed the
corresponding $T_t$-\astralbody{}s of the nodes $u,v,$ and $z,$
and with the corresponding current range.
The full details of the $\O(\lg\sigma)$-time algorithm are given in \Cref{algo:extractionCounting} of \Cref{appendix:queryAlgos}.

\myenv{To summarize, the variant of {\ext} that we design here uses
$3n\lg\sigma+\smallO(n\lg\sigma)$ bits to support {\pathmed} and {\pathcnt} in
$\O(\lg\sigma)$ time, and {\pathrp} in $\O((1+\kappa)\lg\sigma)$ time.
Compared to the original succinct solution~\cite{DBLP:journals/talg/0001MZ16}
based on tree extraction, our variant uses about $3$ times
the space with a minor slow-down in query time, 
but is easily implementable using bitmaps and BP,
both of which have been studied experimentally (see e.g.~\cite{DBLP:conf/alenex/ArroyueloCNS10} and~\cite{DBLP:books/daglib/0038982} for an extensive review).}

\section{Experimental Results}
\myenv{We now conduct experimental studies on data structures for path queries.}
\begin{table*}
\centering
\ra{1.05}
	\begin{tabular}{p{.03\textwidth}p{.15\textwidth}p{.75\textwidth}@{}}
			\cmidrule[0.25ex]{2-3}
			{} & {Symbol} & \multicolumn{1}{c}{Description}\\
	\cmidrule{2-3}
\parbox[t]{3mm}{\multirow{4}{*}{\rotatebox[origin=c]{90}{ \begin{footnotesize}{\textit{pointer-based}}\end{footnotesize} }}} 
		& {\nv} & {\begin{small}{Na{\"i}ve data structure in \Cref{sec:approaches}}\end{small}}\\
				& {\nvlca} & {\begin{small}{Na{\"i}ve data structure in \Cref{sec:approaches}, 
		augmented with $\O(1)$ query-time $LCA$ of~\cite{DBLP:journals/jal/BenderFPSS05}}\end{small}}\\
            & {\extptr} & {\begin{small}{A solution based on tree extraction~\cite{DBLP:journals/talg/0001MZ16} in \Cref{section:dcc:ptr}}\end{small}}\\
            & {\wthpdptr} & {\begin{small}{A non-succinct version of the wavelet tree- and heavy-path decomposition-based solution of~\cite{DBLP:journals/jda/PatilST12} in \Cref{sec:compactRepr}.}\end{small}}\\
	\cmidrule{2-3}
	 & {\nvsuc} & {\begin{small}{Na{\"i}ve data structure of \Cref{sec:approaches}, using succinct data structures to represent the tree structure and weights}\end{small}}\\
    \parbox[t]{3mm}{\multirow{5}{*}{\rotatebox[origin=c]{90}{ \begin{footnotesize}{\textit{succinct}}\end{footnotesize} }}} 
     & {\extrrr} & {\begin{small}{$3n\lg\sigma+\smallO(n\lg\sigma)$-bits-of-space scheme for tree extraction of \Cref{par:threelog}, with compressed bitmaps}\end{small}}\\
     & {\extun} & {\begin{small}{$3n\lg\sigma+\smallO(n\lg\sigma)$-bits-of-space scheme for tree extraction of \Cref{par:threelog}, with uncompressed bitmaps }\end{small}}\\
     & {\wthpdrrr} & {{\setlength{\parindent}{0cm}{\begin{small}{Succinct version of {\wthpd}, with compressed bitmaps}\end{small}}}}\\
     & {\wthpdun} & {{\setlength{\parindent}{0cm}{\begin{small}{Succinct version of {\wthpd}, with uncompressed bitmaps}\end{small}}}}\\
	\cmidrule[0.25ex]{2-3}
	\end{tabular}
\caption{The implemented data structures and the abbreviations used to refer to them.}
\label{tables:tabAbbr}
\end{table*}

\subsection{Implementation}\label{sec:implementation}\label{sec:approaches}
For ease of reference, we outline the data structures implemented in \Cref{tables:tabAbbr}.

Na{\"i}ve approaches (both plain pointer-based
{\nv/\nvlca} and succinct {\nvsucc}) resolve a query on the path $P_{x,y}$ by explicitly traversing it from $x$ to $y$.
At each encountered node, we either (i) collect its weight into an array (for \pathmed);
(ii) check if its weight is in the query range (for \pathcnt); (iii) if the check
 in (ii) succeeds, we collect the node into a container (for \pathrp).
In \pathmed, we subsequently call a standard {\textit{introspective selection}} algorithm~\cite{DBLP:journals/spe/Musser97} over the array of collected weights.
Depths and parent pointers, explicitly stored at each node,
guide in upwards traversal from $x$ and $y$ to their common ancestor. 
Plain pointer-based tree topologies are stored using \textit{forward-star}~\cite{DBLP:books/daglib/0069809} representation.
In \nvlca, we equip \nv{} with the linear-space and $\O(1)$-time LCA-support structure of~\cite{DBLP:journals/jal/BenderFPSS05}.

Succinct structures 
{\extrrr/\extun/\wthpdrrr/\wthpdun} are implemented with the help
of the succinct data structures library \sdslite{} of Gog et al.~\cite{DBLP:conf/wea/GogBMP14}.
To implement {\wthpd} and the practical variant of {\ext} we designed in \Cref{par:threelog},
two types of bitmaps are used: a compressed bitmap~\cite{DBLP:journals/talg/RamanRS07}
(implemented in {\classname{sdsl::rrr\_vector}} of \sdslite)
and plain bitmap (implemented in {\classname{sdsl::bit\_vector}} of \sdslite).
For {\nvsuc}, the weights are stored using $\ceil{\lg\sigma}$ bits each in a sequence
and the structure theoretically occupies $2n+n\lg\sigma + \smallO(n\lg\sigma)$ bits.
For uniformity, across our data structures, tree navigation is provided solely by a BP representation
based on~\cite{DBLP:journals/talg/GearyRR06} (implemented in {\classname{sdsl::bp\_support\_gg}}),
chosen on the basis of our benchmarks.

Plain pointer-based implementation {\extptr} is an implementation
of the solution by He et al.~\cite{DBLP:journals/talg/0001MZ16}
for the pointer-machine model, which uses tree extraction.
In it, the \astralbody{}s $x_0 \in T_0,\,x_1 \in T_1$ for
each node that arises in the hierarchy of extractions, as well as the depths in $T,$
are explicitly stored. Similarly, {\wthpdptr} is a plain pointer-based implementation
of the data structure by Patil et al.~\cite{DBLP:journals/jda/PatilST12}.
The relevant source code is accessible at~\url{https://github.com/serkazi/tree_path_queries}.

\subsection{Experimental setup}\label{sec:results}
\begin{table*}
\centering
\resizebox{\columnwidth}{!}{%
\ra{1.3}
\rowcolors{1}{}{lightgray}
\begin{tabular}{@{}lrrrrrp{5cm}@{}}
\toprule
{} & \multicolumn{1}{c}{\texttt{num nodes}} & \multicolumn{1}{c}{\texttt{diameter}} & \multicolumn{1}{c}{\texttt{$\sigma$}} & \multicolumn{1}{c}{\texttt{$\log{\sigma}$}} & \multicolumn{1}{c}{\texttt{$H_0$}} & {Description}\\
\cmidrule{1-7}
{\euosm} & {{\texttt{27,024,535}}} & {{\texttt{109,251}}} & {{\texttt{121,270}}} & {{\texttt{16.89}}} & {{\texttt{9.52}}} & {An MST we constructed over map of Europe~\cite{OpenStreetMap}}\\
{\eudimacs} & {{\texttt{18,010,173}}} & {{\texttt{115,920}}} & {{\texttt{843,781}}} & {{\texttt{19.69}}} & {{\texttt{8.93}}} & {An MST we constructured over European road network~\cite{kitdimacs}}\\
{\euemst} & {{\texttt{50,000,000}}} & {{\texttt{175,518}}} & {{\texttt{5020}}} & {{\texttt{12.29}}} & {{\texttt{9.95}}} & {An Euclidean MST we constructed over DEM of Europe~\cite{srtm}}\\
{\mars} & {{\texttt{30,000,000}}} & {{\texttt{164,482}}} & {{\texttt{29,367}}} & {{\texttt{14.84}}} & {{\texttt{13.23}}} & {An Euclidean MST we constructed over DEM of Mars~\cite{marsnasa}}\\
\bottomrule
\end{tabular}
}
\caption{
    Datasets metadata. 
    DEM stands for Digital Elevation Model, and MST for minimum spanning tree.
    Weights are over $\{0,1,\ldots,\sigma-1\},$ and $H_0$ is the entropy of the multiset of weights.
    In DEM, elevation (in meters) is used as weights. For \euosm{}, distance in meters between
    locations, 
    and for \eudimacs, travel time between locations, 
    for a proprietary ``car'' profile in tenths of a second, are used as weights.}
\label{tables:datasetTable}
\end{table*}

The platform
used is a $128$GiB RAM, Intel(R) Xeon(R) Gold $6234$ CPU $3.30$GHz server running 
4.15.0-54-generic 58-Ubuntu SMP x86\_64 kernel. 
The build is due to {\texttt{clang-8}} with {\texttt{-g,-O2,}}\\
{\texttt{-std=c++17,mcmodel=large,-NDEBUG}} flags.
Our datasets originate from geographical information systems (GIS).
In \Cref{tables:datasetTable}, the relevant meta-data on our datasets is given.

We generated query paths
by choosing a pair uniformly at random (u.a.r.).
To generate a range of weights, $[a,b],$ we follow
the methodology of~\cite{DBLP:conf/spire/ClaudeMN10}
and consider {\texttt{large}}, {\texttt{medium}}, and {\texttt{small}} configurations:
given $K,$ we generate the left bound $a \in [W]$ 
u.a.r., whereas $b$ 
is generated u.a.r. from $[a,a+\ceil{\frac{W-a}{K}}].$
We set $K = 1,10,$ and $100$ for respectively {\texttt{large}}, {\texttt{medium}}, and {\texttt{small}}.
To counteract skew in weight distribution in some of the datasets, when generating the weight-range $[a,b]$,
we in fact generate a pair from $[n]$ rather than $[\sigma]$ and map the positions
to the sorted list of input-weights, ensuring the number of nodes covered by the generated weight-range
to be proportional to $K^{-1}.$ 

\subsection{Space performance and construction costs}
A single data structure we implement (be it ever \nv-, \ext-, or \wthpd-family),
taken individually, answers all three types of queries (\pathmed, \pathcnt, and \pathrp).
Hence, we consider space consumption first.

\newcommand{\moj}[2]{\textbf{#1}/\texttt{#2}}
\begin{table*}
\begin{small}
\centering
\resizebox{\columnwidth}{!}{%
\ra{1.05}
    \begin{tabular}{@{}ll*{9}{r}@{}}
    \toprule
    {} & {\multirow{1}{*}{Dataset}} & \multirow{1}{*}{\nv} & \multirow{1}{*}{\nvlca} & \multirow{1}{*}{\wthpdptr} & \multirow{1}{*}{\extptr} & \multirow{1}{*}{\nvsuc} & 
        \multicolumn{1}{c}{\extrrr} & \multicolumn{1}{c}{\extun}  & \multicolumn{1}{c}{\wthpdrrr} & \multicolumn{1}{c}{\wthpdun}\\
    \midrule
	\parbox[t]{2mm}{\multirow{4}{*}{\rotatebox[origin=c]{90}{\texttt{space}}}}
	& {\euosm}  & {\laski{406.3}}& {\laski{972.1}} & {\laski{3801}} & {\laski{5943}} & {\laski{21.71}}&{\laski{59.85}}&{\laski{75.74}}&{\laski{21.71}}&{\laski{34.42}}\\
	& {\eudimacs} & {\laski{406.4}}& {\laski{974.0}} & {\laski{4274}} & {\laski{6768}} & {\laski{34.46}}&{\laski{82.16}}&{\laski{106.0}}&{\laski{29.69}}&{\laski{48.77}}\\
    & {\euemst} & {\laski{394.1}}& {\laski{988.5}} & {\laski{3342}} &{\laski{4613}} & {\laski{19.64}} & {\laski{45.41}}&{\laski{59.15}}&{\laski{19.64}}& {\laski{31.66}}\\
	& {\mars}  & {\laski{386.7}}& {\laski{1005}} & {\laski{3579}} &{\laski{5383}} & {\laski{17.35}} & {\laski{51.71}} & {\laski{66.02}} & {\laski{17.35}} & {\laski{28.80}}\\

    \midrule
	\parbox[t]{2mm}{\multirow{4}{*}{\rotatebox[origin=c]{90}{\texttt{peak/time}}}}
    & {\euosm} &{\moj{491.0}{1}}&{\moj{987.9}{5}}&{\moj{3785}{28}}  &{\moj{9586}{47}}&{\moj{21.71}{1}}&{\moj{295.0}{23}}&{\moj{295.0}{23}}&{\moj{1347}{62}}&{\moj{1347}{61}}\\
    & {\eudimacs} &{\moj{439.8}{1}}&{\moj{1002}{4}}&{\moj{4403}{19}}&{\moj{12382}{37}}&{\moj{29.69}{1}}&{\moj{399.7}{18}}&{\moj{399.7}{18}}&{\moj{1360}{42}}&{\moj{1360}{42}}\\
    & {\euemst} &{\moj{401.0}{2}}&{\moj{1021}{10}}&{\moj{3460}{47}}&{\moj{5286}{67}}&{\moj{19.64}{1}} & {\moj{287.6}{32}}&{\moj{287.6}{32}}&{\moj{1333}{115}}& {\moj{1333}{115}}\\
    & {\mars} &{\moj{392.4}{1}}& {\moj{1016}{5}}&{\moj{3719}{30}} &{\moj{6027}{46}} & {\moj{17.35}{1}}&{\moj{269.3}{22}}&{\moj{269.3}{22}}&{\moj{1337}{69}}&{\moj{1337}{69}}\\
\bottomrule
\end{tabular}
}
\caption{ (upper) Space occupancy of our data structures, in bits per node, when loaded into memory; (lower) peak memory usage (\textbf{m} in bits per node) during construction
and construction time ($t$ in seconds) shown as $\mathbf{m}/t$.}
\label{tables:spaceAll}
\end{small}
\end{table*}

The upper part of the \Cref{tables:spaceAll} shows the space usage of our data structures.
The structures {\nv/\nvlca} are lighter than {\extptr/\wthpdptr}, as expected.
Adding fast $LCA$ support doubles the space requirement for \nv, whereas
succinctness (\nvsucc) uses up to $20$ times less space than \nv.
The difference between {\extptr} and {\wthpdptr}, in turn, is in explicit
storage of the $0$-\astralbody{}s for each of the $\Theta(n\lg\sigma)$ nodes occurring during tree extraction.
In {\wthpdptr}, by contrast, {$\mathtt{rank}_0$} is induced from $\mathtt{rank}_1$ (via subtraction) -- hence
the difference in the empirical sizes of the otherwise $\Theta(n\lg\sigma)$-word data structures.

The succinct {\nvsucc}'s empirical space occupancy is close to the information-theoretic minimum given by $\lg\sigma+2$ (\Cref{tables:datasetTable}).
The structures {\extrrr/\extun} occupy about three times as much, which is consistent with the design of our practical
solution (\Cref{par:threelog}). It is interesting to note that the data structure {\wthpdrrr} 
occupies space close to bare succinct storage of the input alone (\nvsucc).
Entropy-compression significantly impacts both families of succinct
structures, {\wthpd} and {\ext}, saving up to $20$ bits per node when switching from plain bitmap to a compressed one.
Compared to pointer-based solutions (\nv/\nvlca/\wthpdptr/\extptr), we note that {\extrrr/\extun/\wthpdrrr/\wthpdun}
still allow usual navigational operations on $T$, whereas the former shed this redundancy,
to save space, after preprocessing.

Overall, the succinct {\wthpdun/\wthpdrrr/\extun/\extrrr} perform very well, being all well-under $1$ gigabyte
for the large datasets we use. This suggests scalability: when trees are so large as not to fit into main memory,
it is clear that the succinct solutions are the method of choice. 

The lower part in \Cref{tables:spaceAll} shows peak memory usage ({\textbf{m}}, in bits per node) and construction time (t, in seconds), as $\mathbf{m}/t.$
The structures {\extun/\extrrr} are about three times faster than {\wthpdun/\wthpdrrr} to build,
and use four times less space at peak. This is expected, as {\wthpd} builds two different structures (HPD and then WT). 
This is reversed for {\extptr/\wthpdptr};
time-wise, as {\extptr} performs more memory allocations during construction
(although our succinct structures are flattened into a heap layout, 
{\extptr} stores pointers to $T_0/T_1$;
this is less of a concern for {\wthpdptr}, whose very purpose is tree linearisation).

\begin{table*}
\begin{small}
\centering
\resizebox{0.87\columnwidth}{!}{%
\ra{1.05}
    \begin{tabular}{@{}ll*{10}{r}@{}}
    \toprule
    & {Dataset} & \multirow{1}{*}{\nv} & \multirow{1}{*}{\nvlca} & \multirow{1}{*}{\extptr} & \multirow{1}{*}{\wthpdptr} & \multirow{1}{*}{\nvsuc} & \multicolumn{1}{c}{\extrrr} & \multicolumn{1}{c}{\extun} & \multicolumn{1}{c}{\wthpdrrr} & \multicolumn{1}{c}{\wthpdun}\\
    \midrule
    \parbox[t]{2mm}{\multirow{4}{*}{\rotatebox[origin=c]{90}{\texttt{median}}}} & 
	{\euosm} & {\laski{658}} & {\laski{475}} & {\laski{4.22}} & {\laski{6.10}} & {\laski{7078}} & {\laski{85.3}} & {\laski{51.1}} & {\laski{111}} & {\laski{51.2}}\\ 
	& {\eudimacs} & {\laski{566}} & {\laski{412}} & {\laski{5.16}} & {\laski{6.28}} & {\laski{6556}} & {\laski{84.6}} & {\laski{54.8}} & {\laski{120}} & {\laski{54.7}}\\ 
	& {\euemst} & {\laski{710}} & {\laski{436}} & {\laski{4.44}} & {\laski{5.10}} & {\laski{9404}} & {\laski{106}} & {\laski{81.9}} & {\laski{96.7}} & {\laski{54.9}}\\ 
	& {\mars} & {\laski{472}} & {\laski{298}} & {\laski{4.93}} & {\laski{4.53}} & {\laski{7018}} & {\laski{124}} & {\laski{97.0}} & {\laski{88.3}} & {\laski{49.5}}\\ 
    \midrule
    \parbox[t]{2mm}{\multirow{12}{*}{\rotatebox[origin=c]{90}{\texttt{counting}}}} & 
     {\euosm} & {\laski{238}} & {\laski{140}} & {\laski{6.88}} & {\laski{18.4}} & {\laski{3553}} & {\laski{247}} & {\laski{167}} & {\laski{139}} & {\laski{56.9}} & \parbox[t]{2mm}{\multirow{4}{*}{\rotatebox[origin=c]{90}{\texttt{large}}}}\\ 
        & {\eudimacs} & {\laski{204}} & {\laski{121}} & {\laski{7.31}} & {\laski{19.7}} & {\laski{3300}} & {\laski{253}} & {\laski{178}} & {\laski{142}} & {\laski{57.3}} &\\ 
        & {\euemst} & {\laski{338}} & {\laski{195}} & {\laski{5.97}} & {\laski{11.5}} & {\laski{4835}} & {\laski{215}} & {\laski{168}} & {\laski{105}} & {\laski{55.9}} &\\ 
        & {\mars} & {\laski{232}} & {\laski{174}} & {\laski{5.25}} & {\laski{8.40}} & {\laski{3614}} & {\laski{206}} & {\laski{164}} & {\laski{91}} & {\laski{49.3}} &\\ 
        \cmidrule{2-11}
        & {\euosm} & {\laski{244}} & {\laski{143}} & {\laski{5.47}} & {\laski{17.8}} & {\laski{3555}} & {\laski{213}} & {\laski{146}} & {\laski{129}} & {\laski{54.2}} & \parbox[t]{2mm}{\multirow{4}{*}{\rotatebox[origin=c]{90}{\texttt{medium}}}}\\ 
        & {\eudimacs} & {\laski{209}} & {\laski{124}} & {\laski{6.94}} & {\laski{18.4}} & {\laski{3297}} & {\laski{224}} & {\laski{160}} & {\laski{133}} & {\laski{56.5}} & \\ 
        & {\euemst} & {\laski{339}} & {\laski{195}} & {\laski{4.55}} & {\laski{10.0}} & {\laski{4840}} & {\laski{178}} & {\laski{140}} & {\laski{100}} & {\laski{54.9}} & \\ 
        & {\mars} & {\laski{237}} & {\laski{143}} & {\laski{5.91}} & {\laski{8.74}} & {\laski{3613}} & {\laski{199}} & {\laski{154}} & {\laski{89.7}} & {\laski{48.9}} & \\ 
        \cmidrule{2-11}
        & {\euosm} & {\laski{239}} & {\laski{139}} & {\laski{5.25}} & {\laski{15.4}} & {\laski{3551}} & {\laski{190}} & {\laski{132}} & {\laski{119}} & {\laski{53.9}} & \parbox[t]{2mm}{\multirow{4}{*}{\rotatebox[origin=c]{90}{\texttt{small}}}}\\ 
        & {\eudimacs} & {\laski{209}} & {\laski{123}} & {\laski{5.25}} & {\laski{18.9}} & {\laski{3300}} & {\laski{206}} & {\laski{148}} & {\laski{126}} & {\laski{55.2}} & \\ 
        & {\euemst} & {\laski{347}} & {\laski{200}} & {\laski{3.92}} & {\laski{9.34}} & {\laski{4832}} & {\laski{154}} & {\laski{124}} & {\laski{94.9}} & {\laski{53.2}} & \\ 
        & {\mars} & {\laski{238}} & {\laski{144}} & {\laski{4.82}} & {\laski{7.41}} & {\laski{3615}} & {\laski{178}} & {\laski{133}} & {\laski{84.2}} & {\laski{47.6}} & \\ 
		\bottomrule
\end{tabular}
}
\caption{
    Average time to answer a query, from a fixed set of $10^6$ randomly generated path median
    and path counting queries, in microseconds.
    Path counting queries are given in {\texttt{large}}, {\texttt{medium}}, and {\texttt{small}} configurations.
}
\label{tables:countingAndReportingTimes}
\end{small}
\end{table*}

\subsection{Path median queries}
The upper section of \Cref{tables:countingAndReportingTimes} records the mean time
for a single median query (in \microsec) averaged over a fixed set of $10^6$ randomly generated queries.

Succinct structures {\wthpdrrr/\wthpdun/\extrrr/\extun} perform well
on these queries, with a slow-down of at most $20$-$30$ times from their respective pointer-based counterparts.
Using entropy-compression degrades the speed of {\wthpd} almost twice. 
Overall, the families {\wthpd} and {\ext} seem to perform at the same order of magnitude.
This is surprising, as in theory {\wthpd} should be a factor of ${\lg{n}}$ slower.
The discrepancy is explained partly by small average number of segments in HPD,
averaging $9\pm{}2$ for our queries. 
(The number of unary-degree nodes in our datasets is $35\%$-$56\%,$
which makes smaller number of heavy-path segments prevalent.
\myenv{We did not use trees with few unary-degree nodes in our experiments,
as the height of such trees are not large enough to make constructing data structures
for path queries worthwhile.})
When the queries are partitioned by the number of chains in the HPD,
the curves for {\extrrr/\extun} stay flat whereas those for {\wthpdrrr/\wthpdun}
grow linearly (see \Cref{fig:avgVersusHpd} in \Cref{sec:timeVsHdp}). 
Take {\eudimacs} as an example. When the query path is partitioned into $9$ chains,
{\extun} is only slightly faster than {\wthpdun}, but when the query path
contains $19$ chains, {\extun} is about $2.3$ times so.
This suggests to favour the {\ext} family
over {\wthpd} whenever performance in the worst case is important.
Furthermore, navigational operations
in {\extun/\extrrr} and {\wthpdun/\wthpdrrr}, despite of similar theoretical
worst-case guarantees, involve different patterns of using
the {\texttt{rank/select}} primitives. For one, {\wthpdun/\wthpdrrr} does not
call $LCA$ during the search -- mapping of the search ranges when descending down the recursion is accomplished
by a single {\texttt{rank}} call, whereas {\extun/\extrrr}
computes $LCA$ at each level of descent (for its its own
analog of {\texttt{rank}} -- the {\astralbody{}} computation in \Cref{algo:algo02}).
Now, $LCA$ is a non-trivial combination of {\texttt{rank/select}} calls.
The difference between {\extun/\extrrr} and {\wthpdun/\wthpdrrr}
will therefore become pronounced in a large enough tree; with tangible HPD sizes, the constants
involved in (albeit theoretically $\O(1)$) $LCA$ calls are overcome by $\lg{n}.$

Na{\"i}ve structures {\nv/\nvlca/\nvsucc} are visibly slower in \pathmed{}
than in \pathcnt{} (considered in \Cref{sec:resultsCounting}),
as expected --- for \pathmed, having collected the nodes encountered,
we also call a selection algorithm. In \pathcnt{}, by contrast,
neither insertions into a container nor a subsequent search for median are involved.
Navigation and weights-uncompression in {\nvsucc} render it about $10$ times slower than its plain counterpart.
The {\nvlca} being little less than twice faster than its $LCA$-devoid counterpart, \nv,
is explained by the latter effectively traversing the query path
twice --- once to locate the $LCA,$ and again to answer the query proper.
\myenv{Any succinct solution is about $4$-$8$ times faster than the fastest na{\"i}ve, \nvlca.}

\subsection{Path counting queries}\label{sec:resultsCounting}
The lower section in \Cref{tables:countingAndReportingTimes} records the mean time
for a single counting query (in \microsec) averaged over a fixed set of $10^6$ randomly generated queries,
for {\texttt{large}}, {\texttt{medium}}, and {\texttt{small}} setups.

Structures {\nv/\nvlca/\nvsucc} are insensitive
to $\kappa,$ as the bottleneck is in physically traversing the path. 

Succinct structures {\wthpdun/\wthpdrrr} and {\extun/\extrrr} 
exhibit decreasing running times as one moves from {\texttt{large}} to {\texttt{small}} ---
as the query weight-range shrinks, so does the chance
of branching during the traversal of the implicit range tree.
The fastest (uncompressed) {\wthpdun} and the slowest (compressed) {\extrrr} succinct solutions differ by a factor of $4,$
which is intrinsically larger constants in {\extrrr}'s implementation compounded
with slower {\texttt{rank/select}} primitives in compressed bitmaps, at play.
\myenv{The uncompressed {\wthpdun} is about $2$-$3$ times faster than {\extun}, the gap narrowing
towards the {\texttt{small}} setup.}
The slowest succinct structure, {\extrrr}, is nonetheless competitive with the {\nv/\nvlca} already in {\texttt{large}} configuration,
with the advantage of being insensitive to tree topology.

In {\extptr}-{\wthpdptr} pair, \wthpdptr{} is $2$-$3$ times slower.
This is predictable, as the inherent $\lg{n}$-factor
slow-down in {\wthpdptr} is no longer offset by differing memory access patterns -- following a pointer ``downwards''
(i.e. $0/1$-{\astralbody} in {\extptr} and {$\mathtt{rank}_{0/1}()$}
in {\wthpdptr}) each require a single memory access.

\begin{table*}
    \begin{small}
\centering
\ra{1.00}
    \begin{tabular}{@{}lll*{9}{r}r@{}}
    \toprule
        &\multirow{1}{*}{Dataset} & \multirow{1}{*}{$\kappa$} & \multirow{1}{*}{\nv} & \multirow{1}{*}{\nvlca} & \multirow{1}{*}{\extptr} & \multirow{1}{*}{\wthpdptr} & \multirow{1}{*}{\nvsuc} & \multirow{1}{*}{\extrrr} & \multirow{1}{*}{\extun} & \multirow{1}{*}{\wthpdrrr} & \multirow{1}{*}{\wthpdun} & \\
    \midrule
        & {\euosm} & {\laski{9,840}} & {\laski{356}} & {\laski{256}} & {\laski{184}} & {\laski{70.7}} & {\laski{3766}} & \hatched{1}{4} & {\parbox[t]{2mm}{\multirow{4}{*}{\rotatebox[origin=c]{90}{\texttt{large}}}}} \\
        & {\eudimacs} & {\laski{9,163}} & {\laski{309}} & {\laski{224}} & {\laski{147}} & {\laski{66.8}} & {\laski{3485}} & \hatched{2}{4} & {} \\
        & {\euemst} & {\laski{14,211}} & {\laski{389}} & {\laski{241}} & {\laski{140}} & {\laski{77.5}} & {\laski{4926}}  & \hatched{3}{4} & {} \\
        & {\mars} & {\laski{10,576}} & {\laski{267}} & {\laski{178}} & {\laski{89.2}} & {\laski{55.1}} & {\laski{3668}} &     \hatched{4}{4} & {} \\
   \cmidrule{1-8}
        & {\euosm} & {\laski{1,093}} & {\laski{322}} & {\laski{222}} & {\laski{43.7}} & {\laski{28.8}} & {\laski{3706}} & \hatched{5}{4} & {\parbox[t]{2mm}{\multirow{4}{*}{\rotatebox[origin=c]{90}{\texttt{medium}}}}} \\
        & {\eudimacs} & {\laski{1,090}} & {\laski{277}} & {\laski{196}} & {\laski{34.0}} & {\laski{29.7}} & {\laski{3434}} & \hatched{6}{4} & {} \\ 
        & {\euemst} & {\laski{1,464}} & {\laski{354}} & {\laski{206}} & {\laski{32.1}} & {\laski{20.1}} & {\laski{4880}} & \hatched{7}{4} & {} \\
        & {\mars} & {\laski{1,392}} & {\laski{250}} & {\laski{151}} & {\laski{22.1}} & {\laski{15.6}} & {\laski{3639}} & \hatched{8}{4} & {} \\
   \cmidrule{1-12}
        & {\euosm} & {\laski{182}} & {\laski{311}} & {\laski{212}} & {\laski{13.8}} & {\laski{19.0}} & {\laski{3685}} & {\laski{1965}} & {\laski{485}} & {\laski{795}} & {\laski{226}} & \parbox[t]{2mm}{\multirow{4}{*}{\rotatebox[origin=c]{90}{\texttt{small}}}} \\
        & {\eudimacs} & {\laski{236}} & {\laski{271}} & {\laski{193}} & {\laski{13.2}} & {\laski{21.0}} & {\laski{3529}} & {\laski{2518}} & {\laski{632}} & {\laski{1043}} & {\laski{292}} & \\ 
        & {\euemst} & {\laski{215}} & {\laski{353}} & {\laski{203}} & {\laski{10.2}} & {\laski{12.7}} & {\laski{4873}} & {\laski{1276}} & {\laski{378}} & {\laski{590}} & {\laski{205}} & \\
        & {\mars} & {\laski{117}} & {\laski{242}} & {\laski{145}} & {\laski{8.88}} & {\laski{9.57}} & {\laski{3632}} & {\laski{881}} & {\laski{278}} & {\laski{475}} & {\laski{162}} & \\
   \bottomrule
\end{tabular}
\caption{Average time to answer a path reporting query, from a fixed set of $10^6$ randomly generated path
    reporting queries, in microseconds. The queries are given
    in {\texttt{large}}, {\texttt{medium}}, and {\texttt{small}} configurations. Average output size
    for each group is given in column $\kappa.$}
\label{tables:reportingTimes}
\HatchedCell{start1}{end1}{%
  pattern color=black!70,pattern=north east lines}
\HatchedCell{start2}{end2}{%
  pattern color=black!70,pattern=north west lines}
\HatchedCell{start3}{end3}{%
  pattern color=black!70,pattern=north east lines}
\HatchedCell{start4}{end4}{%
  pattern color=black!70,pattern=north west lines}
\HatchedCell{start5}{end5}{%
  pattern color=black!70,pattern=north east lines}
\HatchedCell{start6}{end6}{%
  pattern color=black!70,pattern=north west lines}
\HatchedCell{start7}{end7}{%
  pattern color=black!70,pattern=north east lines}
\HatchedCell{start8}{end8}{%
  pattern color=black!70,pattern=north west lines}
    \end{small}
\end{table*}

\subsection{Path reporting queries}
\Cref{tables:reportingTimes} records the mean time
for a single reporting query (in \microsec) averaged over a fixed set of $10^6$ randomly generated queries,
for {\texttt{large}}, {\texttt{medium}}, and {\texttt{small}} setups.

Structures {\wthpdrrr/\wthpdun/\extrrr/\extun} 
recover each reported node's weight in $\O(\lg\sigma)$ time.
Thus, when $\lg{n} \ll \kappa$, the query time for both {\ext} and {\wthpd} families become $\O(\kappa\cdot{}\log\sigma).$ 
(At this juncture, a caveat is in order:
design of {\wthpd}'s in \Cref{sec:HPD} allows
a \pathrp-query to only return the {\textit{index in the
array $C$}} --- not the original preorder identifier of the node, as does the {\ext}.)
When $\kappa$ is large, therefore,
these structures are not suitable for use in \pathrp, as \nv/\nvlca/\nvsucc are clearly superior
($\O((1+\kappa)\lg{n})$ {\textit{vs}} $\O(\kappa)$),
and we confine the experiments
for {\extrrr/\extun/\wthpdrrr/\wthpdun} to the {\texttt{small}}
setup only (bottom-right corner in \Cref{tables:reportingTimes}). 

We observe that
the succinct structures {\extun} and {\wthpdun} are competitive with
{\nv/\nvlca}, in {\texttt{small}} setting:
informally, time saved in locating the nodes to report
is used to uncompress the nodes' weights (whereas in {\nv/\nvlca}
the weights are explicit). Between the succinct {\ext} and {\wthpd},
clearly {\wthpd} is faster, as {\texttt{select()}} on a sequence
as we go up the wavelet tree tend to have lower constant factors than the counterpart
operation on BP.

Structures {\wthpdptr} and {\extptr} exhibit same order
of magnitude in query time, with the former being sometimes about $2$ times faster on non-{\texttt{small}} setups.
Among two somewhat intertwined reasons,
one is that {\wthpdptr} returns an index to the permuted array, as noted above.
(Converting to the original id would necessitate an additional memory access.)
Secondly, in the implicit range tree during the $2d$ search in {\wthpdptr},
when the current range is contained within the query interval, we start
reporting the node weights by merely incrementing a counter --- position in the WT sequence.
By contrast, in such situations {\extptr} iterates through the nodes being reported 
calling {\texttt{parent()}} for the current node, 
which is one additional memory access compared to {\wthpdptr} (at the scale of \microsec, this matters). 
Indeed, operations on trees tend to be little more expensive than similar
operations on sequences.

Structures {\nv/\nvlca/\nvsucc} are less sensitive
to the query weight range's magnitude, since they simply scan the path along with pushing into a container. 
The differences in running time in \Cref{tables:reportingTimes}
between the configurations
are thus accounted for by container operations' cost.
Na{\"i}ve structures' query times for \pathrp{} being dependent solely
on the query path's length, they are unfeasible
for large-diameters trees (whereas they may be suitable
for shallow ones, e.g. originating from ``small-world'' networks).

\pgfplotsset{every axis/.append style={
                    label style={font=\ttfamily},
                    y tick label style={font=\small\ttfamily,rotate=90},
                    x tick label style={font=\small\ttfamily}
                    }}
\tikzset{>=stealth}

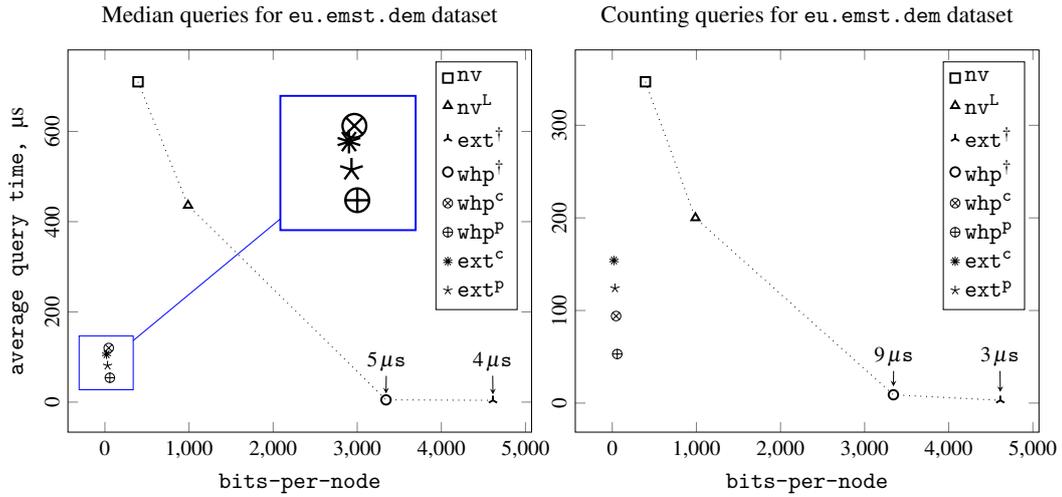
\begin{figure}
\begin{adjustbox}{width=\textwidth}

    \begin{tikzpicture}[spy using outlines={rectangle, lens={scale=2.5}, connect spies}]
        \begin{scope}
    \begin{axis}[
        group style={
            group size=2 by 1,
            ylabels at=edge left
        },
        legend cell align=left,
        legend pos=north east,
        ylabel={average query time, \microsec}, 
        xlabel={bits-per-node},
        title= {Median queries for {\euemst} dataset},
    ]
            \tikzset{spy using outlines={rectangle, lens={scale=3}, connect spies}}
        \addplot [
                scatter/classes={
                a={mark=square,draw=black,thick},
                b={mark=triangle,draw=black,thick},
                c={mark=Mercedes star,draw=black,thick},
                d={mark=o,black,thick},
                e={mark=otimes,draw=black},
                f={mark=oplus,draw=black},
                g={mark=10-pointed star,black},
                h={mark=star,black}
            },
            scatter, only marks,
            scatter src=explicit symbolic,
        ]
        table [meta=label] {
            x   y    label
            394.1 710 a
            988.5 436 b 
            4613 4 c 
            3342 5 d 
            45.41 120 e 
            59.15 54 f 
            19.64 106 g 
            31.66 81  h 
        };

        \node[inner sep=2,outer sep=0] (labelC) at (4608,90) {\small${\mathtt{4\,\mu{}s}}$};
        \coordinate (coordC) at (4613,17);
        \path[->,ultra thin] (labelC) edge[draw] (coordC);

        \node[inner sep=2,outer sep=0] (labelD2) at (3347,90) {\small${\mathtt{5\,\mu{}s}}$};
        \coordinate (coordD2) at (3342,19);
        \path[->,ultra thin] (labelD2) edge[draw] (coordD2);

        \draw[thin,dotted] (axis cs:394,710) -- (axis cs:988,436) -- (axis cs:3342,5) -- (axis cs:4613,4);
        \legend{\nv,\nvlca,\extptr,\wthpdptr,\wthpdrrr,\wthpdun,\extrrr,\extun}
       \coordinate (spypoint2) at (15,87);
        \coordinate (magnifyglass2) at (2890,530);
  \end{axis}
  \spy [blue,size=2cm] on (spypoint2) in node[fill=white] at (magnifyglass2);
  \end{scope}
    \begin{scope}[shift={(7.5,0)}]
    \begin{axis}[
        group style={
            group size=2 by 1,
            ylabels at=edge left
        },
        legend cell align=left,
        legend pos=north east,
        xlabel={bits-per-node},
        title= {Counting queries for {\euemst} dataset}
    ]
        \addplot [
                scatter/classes={
                a={mark=square,draw=black,thick},
                b={mark=triangle,draw=black,thick},
                c={mark=Mercedes star,draw=black,thick},
                d={mark=o,black,thick},
                e={mark=otimes,draw=black},
                f={mark=oplus,draw=black},
                g={mark=10-pointed star,black},
                h={mark=star,black}
            },
            scatter, only marks,
            scatter src=explicit symbolic,
        ]
        table [meta=label] {
            x   y    label
            394.1 347 a
            988.5 200 b
            4613 3 c 
            3342 9 d 
            45.41 94 e 
            59.15 53 f 
            19.64 154 g 
            31.66 124 h 
        };

        \node[inner sep=2, outer sep=0] (labelC) at (4608,50) {\small${\mathtt{3\,\mu{}s}}$};
        \coordinate (coordC) at (4613,12);
        \path[->,ultra thin] (labelC) edge[draw] (coordC);

        \node[inner sep=2,outer sep=0] (labelD2) at (3347,50) {\small${\mathtt{9\,\mu{}s}}$};
        \coordinate (coordD2) at (3342,18);
        \path[->,ultra thin] (labelD2) edge[draw] (coordD2);

        \draw[thin,dotted] (axis cs:394,347) -- (axis cs:988,200) -- (axis cs:3342,9) -- (axis cs:4613,3);
        \legend{\nv,\nvlca,\extptr,\wthpdptr,\wthpdrrr,\wthpdun,\extrrr,\extun}
        \coordinate (spypoint) at (1.7,68);
        \coordinate (magnifyglass) at (15.5,220);
  \end{axis}
        \end{scope}

\end{tikzpicture}
\end{adjustbox}
\caption{Visualization of some of the entries in \Cref{tables:countingAndReportingTimes}. Inner rectangle magnifies
the mutual configuration of the succinct data structures {\wthpdun},{\wthpdrrr},{\extun}, and {\extrrr}. The succinct 
na{\"i}ve structure {\nvsucc} is not shown.} 
\label{fig:fig002}
\end{figure}

\subparagraph{Overall evaluation.}
We visualize in \Cref{fig:fig002} some typical entries in \Cref{tables:countingAndReportingTimes}
to illustrate the structures clustering along the space/time trade-offs:
{\nv/\nvlca} (upper-left corner) are lighter in terms of space, but slow; pointer-based
{\extptr/\wthpdptr} are very fast, but space-heavy.
Between the two extremes of the spectrum,
the succinct structures {\extrrr/\extun/\wthpdrrr/\wthpdun},
whose mutual configuration is shown magnified in inner rectangle,
are space-economical and yet offer fast query times.

\section{Conclusion}\label{sec:conclusion}
We have designed and experimentally evaluated recent algorithmic proposals
in path queries in weighted trees, by either faithfully replicating them
or offering practical alternatives. 
Our data structures include both plain pointer-based and succinct implementations.
Our succinct realizations are themselves further specialized to be either plain or entropy-compressed.

We measure both query time and space performance of our data structures on large practical sets.
We find that the succinct structures we implement offer an attractive alternative
to plain pointer-based solutions, in scenarios with critical space- and query time-performance
and reasonable tolerance to slow-down.
Some of the structures we implement (\wthpdrrr) occupy space equal to bare compressed storage (\nvsuc) of the object
and yet offer fast queries on top of it, while another structure (\extrrr/\extun) occupies space comparable to \nvsuc, offers fast queries
and low peak memory in construction. While {\wthpd} succinct family performs well in average case,
thus offering attractive trade-offs between query time and space occupancy,
{\ext} is robust to the structure of the underlying tree, and is therefore recommended
when strong worst-case guarantees are vital.

Our design of the practical succinct structure based on tree extraction (\ext)
results in a theoretical space occupancy of $3n\lg\sigma+\smallO(n\lg\sigma)$ bits,
which helps explain its somewhat higher empirical space cost when compared to the succinct {\wthpd} family.
At the same time, verbatim implementation of the space-optimal solution 
by He et al.~\cite{DBLP:journals/talg/0001MZ16} draws on 
components that are likely to be cumbersome in practice.
For the path query types considered in this study, therefore,
realization of the theoretically time- and space-optimal data structure --- or indeed some feasible
alternative thereof --- remains an interesting open problem in algorithm engineering.

\bibliography{final_bib}

\appendix

\section{Query Algorithms}\label{appendix:queryAlgos}\label{appendix:queryOverExt}
We enter \Cref{algo:algo00} with several parameters --
the current tree $T$, the query nodes $u,v$, the $LCA\,\,z$ of the two nodes, the quantile $k$ we are looking for,
the weight-range $[a,b],$ and a number $w.$ These are initially set, respectively, to be the outermost tree, the original
query nodes, the $LCA$ of the original query nodes, the median's index (i.e.~half the length of the corresponding
path in the original tree), the weight range $[\sigma],$ and the weight
of the $LCA$ of the original nodes. We maintain the invariant that $T$ is weighted over $[a,b],$
$z$ is the $LCA$ of $u$ and $v$ in $T.$ 
Line~\ref{algo:algo00lineref01} checks whether the current tree is weight-homogeneous. If it is,
we immediately return the current weight $a$ (line~\ref{algo:algo00lineref02}). 
Otherwise, the quantile value we are looking for
is either on the left or on the right half of the weight-range $[a,b].$
In lines~\ref{algo:algo00lineref05}-\ref{algo:algo00lineref11} we check, successively, the ranges $[a_0,b_0]$ and $[a_1,b_1]$ 
to determine how many nodes on the path from $u$ to $v$ in $T$ have weights from the corresponding interval.
The accumulator variable {\texttt{acc}} keeps track of these values and is certain to always be at most $k.$
When the next value of {\texttt{acc}} is about to become larger than $k$ (line~\ref{algo:algo00lineref11}), we are certain
that the current weight-interval is the one we should descend to (line~\ref{algo:algo00lineref12}).
The invariants are maintained in line~\ref{algo:algo00lineref06}: there, we calculate the \astralbody{}s of the current
nodes $u,v,$ and $z$ in the extracted tree we are looking at.

It is clear that $\O(\lg\sigma)$ levels of recursion are explored. At each level of recursion,
a constant number of {\texttt{view\_of()}} and {\texttt{depth()}} operations are performed (lines~\ref{algo:algo00lineref06}-\ref{algo:algo00lineref07}).
Hence, assuming the $\O(1)$-time for the latter operations themselves, we have a $\O(\lg\sigma)$ query-time algorithm, overall.

\begin{small}
\begin{algorithm}[H]
    \caption{Selection: return the $k$-th smallest weight on the path from $u \in T$ to $v \in T$}
    \label{algo:algo00}
    \small
	\begin{algorithmic}[1]
		\Require{$z = LCA(u,v),\,a\leq b,\,\,k \geq 0$}
		\Function{select}{$T,u,v,z,k,w,[a..b]$}
        \If{$a == b$} \label{algo:algo00lineref01}
			\State \Return{$a$} \label{algo:algo00lineref02}
		\EndIf
		\Let{$\mathtt{acc}$}{$\mathtt{0}$}
        \For{$\mathtt{t \in 0..1}$} \label{algo:algo00lineref05}
			\Let{$\mathtt{iu,iv,iz}$}{$\mathtt{view\_of(u,t),\,view\_of(v,t),\,view\_of(z,t)}$} \label{algo:algo00lineref06}
			\Let{$\mathtt{du,dv,dz}$}{$\mathtt{depth(B_t,ix),\,depth(B_t,iy),\,depth(B_t,iz)}$} \label{algo:algo00lineref07}
			\Let{$\mathtt{dw}$}{$\mathtt{du+dv-2\cdot{}dz}$}
			\If{$\mathtt{a_t \leq w \leq b_t}$} \Comment{$[a_0..b_0] = [a..c],\,\,[a_1..b_1] = [c+1..b],\,\,c=(a+b)/2$}
				\Let{$\mathtt{dw}$}{$\mathtt{dw+1}$}
			\EndIf
			\If{$\mathtt{acc+dw > k}$} \label{algo:algo00lineref11}
				\State \Return{\sc{select}($\mathtt{T_t,iu,iv,iz,k-acc,w,[a_t..b_t]}$)} \label{algo:algo00lineref12}
			\EndIf
			\Let{$\mathtt{acc}$}{$\mathtt{acc+dw}$}
		\EndFor
        \Assert{\texttt{false}}; \Comment{unreachable statement -- line~\ref{algo:algo00lineref12} should execute at some point}
		\EndFunction
	\end{algorithmic}%
\end{algorithm}
\end{small}

\Cref{algo:extractionCounting} is adapted from~\cite{DBLP:journals/talg/0001MZ16}, and reasoning similar
to \Cref{algo:algo00} applies. Now we have a weight-range $[p,q],$
and maintain that $[p,q] \cap{} [a,b] \neq \emptyset$ 
(the appropriate action is in line~\ref{listingextractionCountinglabel02}). 
In line~\ref{listing:extractionCountinglabel01} we check if the query range $[p,q]$
is completely inside the the current range. If so, we return all the nodes (if $report$ argument
is set to {\sc{True}}) and the number thereof (for counting case).
If not, we descend to $T_0$ and $T_1$ (line~\ref{listing:extractionCountinglabel03}), as discussed previously.
\Cref{algo:extractionCounting} emulates traversal of a path in range tree, maintaining
the current weight range $[a,b]$ and halving at at each step (line~\ref{lines:line14}).
As operations in lines~\ref{lines:line15} and \ref{lines:line16} are constant-time, the algorithm runs in time $\O(\lg{\sigma}).$

\begin{algorithm}[H]
     \caption{Counting and reporting.}
	    \label{algo:extractionCounting}
        \small
		\begin{algorithmic}[1]
   		\Require{$z = LCA(u,v),\,p\leq q$}
		\Function{countreport}{$T,u,v,z,w,[p,q],[a,b],vec=null,report=False$}
            \If{$p \leq a \leq b \leq q$} \label{listing:extractionCountinglabel01}
			\If{$report$}
				\For{$\mathtt{pu} \in \mathcal{A}(u)\,\textrm{and}\,\mathtt{pu \neq z}$}
					\Let{$\mathtt{vec}$}{$\mathtt{vec+original\_node(pu)}$}
				\EndFor
				\For{$\mathtt{pv} \in \mathcal{A}(v)\,\textrm{and}\,\mathtt{pv \neq z}$}
					\Let{$\mathtt{vec}$}{$\mathtt{vec+original\_node(pv)}$}
				\EndFor
				\If{$a \leq w \leq b$}
					\Let{$\mathtt{vec}$}{$\mathtt{vec+original\_node(pv)}$}
				\EndIf
			\EndIf
			\State \Return{$\mathtt{depth(u)+depth(v)-2depth(z)+1_{w \in [p,q]}}$}
		\EndIf
		\If{$[p,q] \cap{} [a,b] = \emptyset$}
            \State \Return{0} \label{listingextractionCountinglabel02}
		\EndIf
		\Let{$\mathtt{res}$}{$\mathtt{0}$}
            \For{$\mathtt{t \in 0..1}$}\label{listing:extractionCountinglabel03} \Comment{$[a_0,b_0] = [a,m],\,\,[a_1,b_1] = [m+1,b],\,\,m=(a+b)/2$} \label{lines:line14}
            \Let{$\mathtt{iu,iv,iz}$}{$\mathtt{view\_of(u,t),\,view\_of(v,t),\,view\_of(z,t)}$} \label{lines:line15}
            \Let{$\mathtt{du,dv,dz}$}{$\mathtt{depth(B_t,ix),\,depth(B_t,iy),\,depth(B_t,iz)}$} \label{lines:line16}
            \Let{$\mathtt{res}$}{$\mathtt{res}$+{\sc{countreport}($\mathtt{T_t,iu,iv,iz,w,[p,q],[a_t,b_t],vec,report}$)}}
		\EndFor
		\State \Return{$\mathtt{res}$}
		\EndFunction
	\end{algorithmic}
\end{algorithm}

\section{Query-Time Performance Controlled for the Length of HPD}\label{sec:timeVsHdp}
\pgfplotsset{every axis/.append style={
                    label style={font=\ttfamily},
                    x tick label style={font=\scriptsize\ttfamily}
                    }}
 
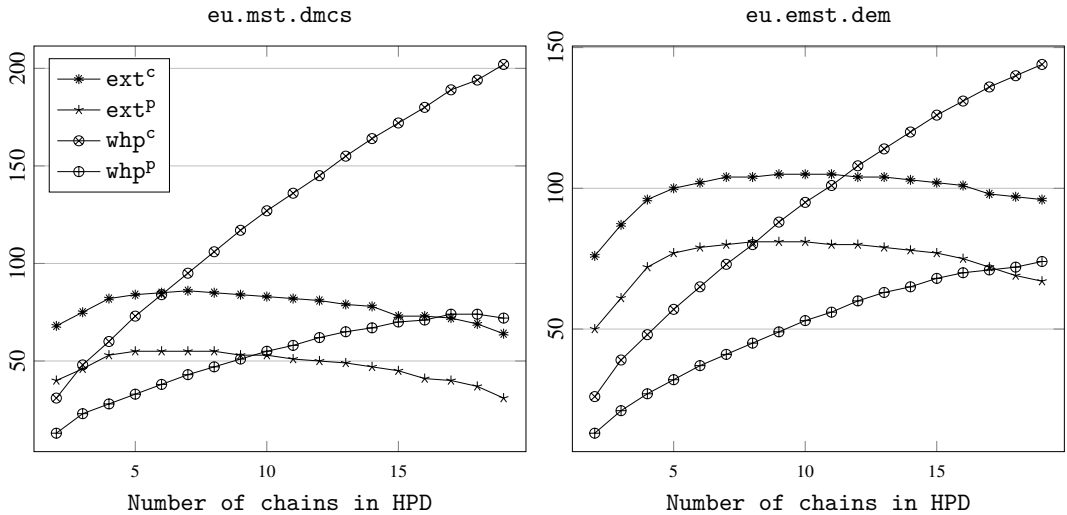
\begin{figure}[H]
    \begin{adjustbox}{width=\textwidth}
    \begin{tikzpicture}
     \begin{scope}
    \begin{axis}[
        enlargelimits=0.05,
        legend pos=north west,
        xticklabel style={text height=1.5ex},
        ymajorgrids,
        xlabel={Number of chains in HPD},
        title style={align=center},
        title = {\eudimacs{}},
    ]
        \addplot[mark=10-pointed star,black] plot coordinates {
            (2,68)
            (3,75)
            (4,82)
            (5,84)
            (6,85)
            (7,86)
            (8,85)
            (9,84)
            (10,83)
            (11,82)
            (12,81)
            (13,79)
            (14,78)
            (15,73)
            (16,73)
            (17,72)
            (18,69)
            (19,64)
        };
        \addplot[mark=star,black] plot coordinates {
            (2,40)
            (3,46)
            (4,53)
            (5,55)
            (6,55)
            (7,55)
            (8,55)
            (9,53)
            (10,53)
            (11,51)
            (12,50)
            (13,49)
            (14,47)
            (15,45)
            (16,41)
            (17,40)
            (18,37)
            (19,31)
        };
        \addplot[mark=otimes,black] plot coordinates {
            (2,31)
            (3,48)
            (4,60)
            (5,73)
            (6,84)
            (7,95)
            (8,106)
            (9,117)
            (10,127)
            (11,136)
            (12,145)
            (13,155)
            (14,164)
            (15,172)
            (16,180)
            (17,189)
            (18,194)
            (19,202)
        };
        \addplot[mark=oplus,black] plot coordinates {
            (2,13)
            (3,23)
            (4,28)
            (5,33)
            (6,38)
            (7,43)
            (8,47)
            (9,51)
            (10,55)
            (11,58)
            (12,62)
            (13,65)
            (14,67)
            (15,70)
            (16,71)
            (17,74)
            (18,74)
            (19,72)
        };
        \legend{\extrrr,\extun,\wthpdrrr,\wthpdun};
    \end{axis}
\end{scope}

\begin{scope}[shift={(7.5,0)}]
    \begin{axis}[
        enlargelimits=0.05,
        legend pos=south east,
        xticklabel style={text height=1.5ex},
        ymajorgrids,
        xlabel={Number of chains in HPD},
        title style={align=center},
         title = {\euemst{}},
    ]
        \addplot[mark=10-pointed star,black] plot coordinates {
            (2, 76)
            (3, 87)
            (4, 96)
            (5, 100)
            (6, 102)
            (7, 104)
            (8, 104)
            (9, 105)
            (10, 105)
            (11, 105)
            (12, 104)
            (13, 104)
            (14, 103)
            (15, 102)
            (16, 101)
            (17, 98)
            (18, 97)
            (19, 96)

        };
        \addplot[mark=star,black] plot coordinates {
                (2, 50)
                (3, 61)
                (4, 72)
                (5, 77)
                (6, 79)
                (7, 80)
                (8, 81)
                (9, 81)
                (10, 81)
                (11, 80)
                (12, 80)
                (13, 79)
                (14, 78)
                (15, 77)
                (16, 75)
                (17, 72)
                (18, 69)
                (19, 67)
        };
        \addplot[mark=otimes,black] plot coordinates {
                (2, 26)
                (3, 39)
                (4, 48)
                (5, 57)
                (6, 65)
                (7, 73)
                (8, 80)
                (9, 88)
                (10, 95)
                (11, 101)
                (12, 108)
                (13, 114)
                (14, 120)
                (15, 126)
                (16, 131)
                (17, 136)
                (18, 140)
                (19, 144)
        };
        \addplot[mark=oplus,black] plot coordinates {
                (2, 13)
                (3, 21)
                (4, 27)
                (5, 32)
                (6, 37)
                (7, 41)
                (8, 45)
                (9, 49)
                (10, 53)
                (11, 56)
                (12, 60)
                (13, 63)
                (14, 65)
                (15, 68)
                (16, 70)
                (17, 71)
                (18, 72)
                (19, 74)
        };
    \end{axis}
\end{scope}
\end{tikzpicture}
\end{adjustbox}
\caption{
   Average time to answer a path median query, controlled for the number of segments in heavy-path decomposition,
   in microseconds. Random fixed query set of size $10^6.$
}
\label{fig:avgVersusHpd}
\end{figure}


\end{document}